\documentclass{amsart}
\usepackage{amssymb}
\usepackage{amsmath}
\usepackage[mathscr]{eucal}
\usepackage{color}
\usepackage[dvips]{graphicx}
\usepackage{enumerate}

\newtheorem{thm}{Theorem}[section]
  
\newtheorem{lem}{Lemma}[section]

\numberwithin{equation}{section}

 \newcommand{\Rthr}{{\mathbb R}^3}
   
      \newcommand{\Zhthr}{{\mathbb Z}_h^3}
       \newcommand{\Ththr}{{\mathbb T}_{\!1/h}^3} 

  \newcommand{\charnh}{{\chi}_{{}_{I_{n,h}} } }
   
   \newcommand{\Fh}{{\mathcal F}_h}
    \newcommand{\iFh}{{\overline{\mathcal F}}_h}
     \newcommand{\Fc}{\mathcal F}
     \newcommand{\iFc}{\overline{\mathcal F}}
 \newcommand{\Dh}{{\mathbb D}_{m,h}}   
   \newcommand{\wDh}{\widehat{\mathbb D}_{m,h}}
    \newcommand{\Dc}{{\mathbb D}_{m} }
 \newcommand{\wDc}{\widehat{\mathbb D}_{m}}

\begin{document}

\title[Discrete Dirac operators on 3D lattices]{On the resolvent convergence of discrete Dirac operators on 3D cubic lattices}

\author{Karl Michael Schmidt}
\address{Karl Michael Schmidt: School of Mathematics,  Cardiff University, Senghennydd Road,
Cardiff CF24 4AG, Wales, UK}
\email{schmidtkm@cardiff.ac.uk}
\author{Tomio Umeda}
\address{Tomio Umeda: Department of
Mathematical Sciences,  University of Hyogo, Himeji 671-2201,
Japan}
\email{umeda@sci.u-hyogo.ac.jp}

\thanks{
Corresponding author: Karl Michael Schmidt\\
T.Umeda was partially supported by
 the Japan Society for the Promotion of Science
     ``Grant-in-Aid for Scientific Research'' (C)
    No.  26400175.
\\
The authors have no competing interests to declare that are relevant to the content of this article. }

\date{}

\begin{abstract}
We prove that the discrete Dirac operators in three dimensions 
converge to the continuum Dirac operators in the strong resolvent sense, but not  
 in the norm
resolvent sense.



\end{abstract}

\maketitle

\section{Introduction}

\noindent
Consider the free Dirac operator in three dimensions,
\begin{eqnarray}
&& \hspace{-35pt}\Dc = - i  \alpha \cdot \nabla + m \beta  \label{eq:Dcont}  \\
& = &\left(\begin{matrix}
 m & 0 & -i\partial_3 &\! -i\partial_1 - \partial_2 \!\\
 0 & m &\! -i\partial_1 + \partial_2 \!& i\partial_3 \\
 -i\partial_3 &\! -i\partial_1 - \partial_2\! & -m & 0 \\
 \!-i\partial_1 + \partial_2\! & i\partial_3 & 0 & -m \end{matrix}\right), \nonumber
 \end{eqnarray}
where $\alpha_1$, $\alpha_2$, $\alpha_3$ and $\beta$ are Dirac matrices
and $m \ge  0$  is the particle mass,
and the corresponding operator discretised on a cubic lattice of mesh size
$h > 0$,
\begin{eqnarray}
 \Dh = \label{eq:Dd}
 \left(\begin{matrix}
 m & 0 & i\partial_{3,h}^* & \!\! i\partial_{1,h}^* \!+ \partial_{2,h}^*\!\\
 0 & m &\! i\partial_{1,h}^* \!- \partial_{2,h}^* \!\!& -i\partial_{3,h}^* \\
 -i\partial_{3,h} &\! -i\partial_{1,h}\! - \partial_{2,h}\!\! & -m & 0 \\
\! -i\partial_{1,h} \!+ \partial_{2,h} \!\!& i\partial_{3,h} & 0 & -m \\
 \end{matrix}\right). 
 \end{eqnarray}
In the latter operator, the partial derivatives are replaced by difference
operators
 \begin{gather*}
 \begin{split}
 & [\partial_{j,h} f](hn)
  := \frac{1}{h}\Big\{ f(h(n+e_j) )-  f(hn)  \Big\},  \\
 & [\partial_{j,h}^* f](hn)
  := \frac{1}{h}\Big\{ f(h(n-e_j)) -  f(hn)  \Big\}  
\end{split}
\end{gather*}
where $ e_j$ is the unit vector in the direction of the $j$th coordinate axis.
We prove that the resolvent of the embedded discrete Dirac operator  $(\Dh \oplus \mathbf 0_h - z)^{-1}$ strongly converges to $(\Dc -z)^{-1}$ in $L^2(\mathbb R^3)^4$ as $h \rightarrow 0$;
see the beginning of Section 2 for details of the embedding. 
The proof presented here in the three dimensional case 
is simpler, shorter and more natural than in
  \cite{SU2023}, where the two dimensional case was discussed.
In addition,
we prove that $(\Dh \oplus \mathbf 0_h - z)^{-1}$ does not converge in the operator norm
sense to $(\Dc -z)^{-1}$ as $h \rightarrow 0$. 
The proof of the non-convergence is an adaptation of the proof of the corresponding result in the
two-dimensional case in \cite{SU2023};
the adaptation enables us to provide an explicit lower estimate of the norm of the resolvent difference as shown in Theorem \ref{thm:nonoco}(ii) below (compare \cite[Theorems 4.11 and 5.7]{CGJ}).
 In order to get norm resolvent convergence,
 one has to modify the discrete Dirac operators in the cubic (resp., square) 
lattices.
Such modifications are discussed in \cite{CGJ} and \cite{N}.
We remark that our imbedding operator $J_h$ defined in the next section 
is different from those introduced in \cite{CGJ} and \cite{N} and exactly preserves the discrete Dirac operator
in $\ell^2(\mathbb Z_h^3)$.

We mention that using central difference operators, instead of 
forward and backward difference operators,  to define $\Dh$ 
does not change the fundamental issue that leads to the positive lower bound for the norm of the resolvent difference in the limit
(Theorem \ref{thm:nonoco}(ii)).
The authors would like to thank John Pryce, Cardiff University, for suggesting the use of central difference operators.

\vspace{8pt}
\section{ Main theorem}

As in \cite{SU2023}, we make the resolvents of $\mathbb D_m$ and of $\mathbb D_{m,h}$ comparable by embedding the discrete Hilbert space $\ell^2(\mathbb Z_h^3)$,
where $\mathbb Z_h^3 = h \mathbb Z^3$,
into the continuum Hilbert space $L^2(\mathbb R^3)$ by extending any function
$f \in \ell^2(\mathbb Z_h^3)$ to the step function
\begin{equation*}
(J_h f)(x) = \sum_{n \in \mathbb Z^3} f(h n)\,\chi_{I_{n,h}}(x) \in L^2(\mathbb  R^3),
\end{equation*}
where $\chi_{I_{n,h}}$ is the characteristic function of the half-open cube
\begin{equation*}
I_{n,h} = \{x \in \mathbb R^3 \mid h n_j \le x_j  <  h\,(n_j+1) \ (j \in \{1,2,3\}) \}.
\end{equation*}
Clearly
$\|J_h f\|_{L^2(\mathbb R^3)}^2 = h^3 \sum_{n \in \mathbb Z^3} |f(hn)|^2$.
The image of the embedding
$L^2(\mathbb Z_h^3) := J_h(\ell^2(\mathbb Z_h^3))$
is a closed subspace of $L^2(\mathbb R^3)$, and $\mathbb D_{m,h}$ naturally
acts on this subspace. Then
$\mathbb D_{m,h} \oplus \mathbf{0}_h$ is the extension of this operator to
all of $L^2(\mathbb R^3)$ by setting it equal to the null operator on the
orthogonal complement of $L^2(\mathbb Z_h^3)$.
The orthogonal projector $P_h$ of $L^2(\mathbb R^3)$ onto $L^2(\mathbb Z_h^3)$
is given by
(cf.\ \cite[eq.~(2.10)]{SU2023})
\begin{equation}\label{eq:ogp}
(P_h \varphi)(x) := \sum_{n \in \mathbb Z^3} \frac 1{h^3} \int_{I_{n,h}} \varphi(y)\,dy\ \chi_{I_{n,h}}(x).
\end{equation}

\medskip

We now state the main theorem, in which
$\mathbf B$ denotes the Banach space of all bounded linear operators in $L^2(\Rthr)^4$, 
equipped with the operator norm. 

\medskip
\begin{thm}\label{thm:nonoco}
Let $z \in \mathbb C \setminus \mathbb R$. Then

\noindent
{\rm (i)} \ $({\mathbb D}_{m,h} \oplus \mathbf 0_h - z )^{-1}$ strongly converges to
$ ({\mathbb D}_{m} - z)^{-1}$ in
$L^2({\mathbb R}^{3})^{4}$ as $h \to 0$.

\vspace{9pt}
\noindent
{\rm (ii)}
\vspace{-24pt}
\begin{align*}
\hspace{-6pt}&\liminf_{h\to 0}
\Vert 
({\mathbb D}_{m,h} \oplus \mathbf 0_h - z )^{-1}  
-  ({\mathbb D}_{m} - z)^{-1}
 \, \Vert_{\mathbf B } 
\ge 
  \max \Big( \frac 1 {|m - z|}, \frac 1 {|m + z|} \Big).
\end{align*}
\end{thm}

\medskip

The proof will use the following asymptotics.

\medskip

\begin{lem}\label{lem:esums}
For $\beta > 0$,
\begin{align}
 \sum_{k\in\mathbb{Z}} e^{-\beta k^2} &= \sqrt{\frac\pi\beta} + O(1) \qquad (\beta\rightarrow 0),
\label{eq:asy1}
\\
 \sum_{k=1}^\infty k^2 \, e^{-\beta k^2} &= \frac{\sqrt\pi}{4 \sqrt\beta^3} + O(\beta^{-1}) \qquad (\beta\rightarrow 0).
\label{eq:asy2}
\end{align}
\end{lem}

\noindent
{\it Proof} of Lemma \ref{lem:esums}.
For the proof of (\ref{eq:asy1}), we split the sum into the parts
with positive and with negative $k$ (these sums are evidently equal) and the
term for $k = 0$, which can be subsumed in the $O(1)$ term.
For a differentiable function $f : [0,\infty) \rightarrow \mathbb C$, integration by parts gives the first-order Euler-Maclaurin formula
\begin{equation*}
\sum_{k=1}^n f(k) = \int_0^n f(t)\,d t + \sum_{k=0}^{n-1} \int_k^{k+1} f'(t)\,(t - k)\,d t
\qquad (n \in \mathbb N).
\end{equation*}
Hence
\begin{equation*}
 \sum_{k=1}^\infty e^{-\beta k^2} 
= \int_0^\infty e^{-\beta t^2}\, d t - \sum_{k=1}^\infty \int_{k-1}^k 2\beta t\,e^{-\beta t^2}\,(t - k)\, d t   
= \frac 1 2 \sqrt{\frac\pi\beta} + O(1),
\end{equation*}
as $|t - k| \le 1$ in each integral in the sum, and also
\begin{align*}
\sum_{k=1}^\infty k^2 e^{-\beta k^2} &= \int_0^\infty t^2\,e^{-\beta t^2}\,d t + \sum_{k=0}^\infty \int_k^{k+1} 2 t\,(1 - \beta t^2)\,e^{-\beta t^2}\,(t - k)\,d t
\\
&= \frac 1{2\beta} \int_0^\infty e^{-\beta t^2}\,d t + O(\beta^{-1})
= \frac{\sqrt\pi}{4\sqrt{\beta}^3} + O(\beta^{-1})
\end{align*}
by an integration by parts in the first integral and as
\begin{equation*}
\left|\sum_{k=0}^\infty \int_k^{k+1} 2 t\,(1 - \beta t^2)\,e^{-\beta t^2}\,(t - k)\,d t \right|
\le \frac 2\beta \int_0^\infty |s|\,|1-s^2|\,e^{-s^2}\,d s.
\hspace{10pt}\blacksquare
\end{equation*}

\medskip

In the proof of Theorem \ref{thm:nonoco}, we  need both the discrete Fourier transform 
$\Fh : L^2(\Zhthr) \to L^2(\Ththr)$ and  
the  Fourier transform $\Fc  : L^2(\Rthr) \to L^2(\Rthr)$,
where 
 \begin{equation*}
 \Ththr= [ -\pi/h, \, \pi/h]^3 
 = \{\xi\in\Rthr \mid  \max \{|\xi_1|, |\xi_2|, |\xi_3|\}\le \pi/h\}.
 \end{equation*}
The precise definitions of $\Fh$, $\Fc$ and  their basic properties 
can be found in   sections 2 and 3 of \cite{SU2023}.
Following  
\cite{SU2023}, we
define the (multiplication) operators $\wDc:= \Fc  \,\mathbb D_m  \iFc$ and 
$\wDh := \iFh D_{m,h} \Fh$.
Since
 \begin{gather}\label{eq:dfourier1}
 \begin{split}
&[ \Fh  (\partial_{j,h}f) ](\xi) = \frac{1}{h}(e^{ih\xi_j} -1)[\Fh f](\xi) \\
&[ \Fh  (\partial_{j,h}^*f) ](\xi) = \frac{1}{h}(e^{-ih\xi_j} -1)[\Fh f](\xi) 
 \end{split}
 \end{gather}
 we can obtain the matrix kernel $\wDh(\xi)$ in an explicit form which shows that
 for each   $\xi \in  \Rthr$, $\big(\wDh(\xi) - \wDc(\xi)\big)\chi_{\Ththr}(\xi) \to 0$ as $h \to 0$. 
 
The orthogonal projection  $Q_{1\!/h}$ of  $L^2(\Rthr)$ 
onto  $L^2(\Ththr)$ is 
defined as the operator of multiplication with the characteristic function of $\Ththr$.
Using the projections $P_h$, 
$Q_{\!1\!/h} \in \mathbf B( L^2(\Rthr))$, we 
extend $\Fh$  and its inverse $\iFh$
 to become elements of  $\mathbf B(L^2(\Rthr))$ 
by setting
\begin{equation}\label{eq:dfourierL2}
\Fh:= Q_{\!1\!/h} \,\Fh P_h,  \quad \iFh:= P_h \iFh Q_{\!1\!/h}.
\end{equation}
It is clear that
\begin{equation}\label{eq:proj-dfour0}
\iFh \Fh = P_h,
\qquad \Fh \iFh = Q_{\!1\!/h},
\end{equation}
and 
\begin{equation}\label{eq:proj-dfour1}
P_h^{\perp} \iFh = \mathbf 0, \quad    Q_{\!1\!/h}^{\perp} \Fh = \mathbf 0.
\end{equation}

\medskip

\medskip
\noindent
{\it Proof} of Theorem \ref{thm:nonoco}.
(i)
Since $\mathcal S (\mathbb R^3)^4$, the Schwartz space of rapidly decreasing functions,
is dense in $L^2(\mathbb R^3)^4$
and
\begin{equation}\label{eq:resest}
\Vert (\Dh-z)^{-1}\oplus \mathbf 0_h  \Vert_{\mathbf B} \le 1/ |\mathfrak{Im}\,z |,
\end{equation}
 it is sufficient to prove that
$$
\lim_{h\to 0}
\Vert \big\{
(\Dh-z)^{-1}\oplus \mathbf 0_h 
- (\Dc -z)^{-1}  
\big\} \varphi \Vert_{[L^2]^4}=0
$$
for 
$\varphi \in \mathcal S (\mathbb R^3)^4$.
 We begin with 
\begin{gather}\label{eq:differnc2}
\begin{split}
\big\{
(\Dh-z)^{-1}\oplus \mathbf 0_h 
&- (\Dc -z)^{-1}  
\big\} \varphi  
- \iFc (\wDc -z)^{-1}\Fc \varphi    \\
&
=
\iFh\big\{
 (\wDh-z)^{-1} \oplus \mathbf 0_h 
\big\} (\Fh - \Fc ) \varphi   \\ 
&
\qquad +
(\iFh - \iFc )\big\{
 (\wDh-z)^{-1} \oplus \mathbf 0_h 
\big\}\Fc \varphi  \\
&\hspace{33pt}
+
\iFc \big\{
 (\wDh-z)^{-1} \oplus \mathbf 0_h   
 - 
(\wDc -z)^{-1}
\big\}\Fc \varphi.   
\end{split}
\end{gather}
\noindent
The $L^2$ norm of the first term  on the right hand side of  (\ref{eq:differnc2})  can  be estimated by
\begin{equation*}
\Vert
 (\wDh-z)^{-1} \oplus \mathbf 0_h 
\Vert_{\mathbf B}
\Vert(\Fh - \Fc)  \varphi \Vert_{L^2(\Rthr)^4}
\le
 \Big(\frac{1}{|\mathfrak{Im}\,z |}\Big) \Vert(\Fh - \Fc)  \varphi \Vert_{L^2(\Rthr)^4}
\end{equation*}
which, by  \cite[Lemma 3.6]{SU2023},
tends to $0$ as $h\to 0$.
The second term  on the right hand side of (\ref{eq:differnc2})  can be written as
\begin{gather}\label{eq:differnc5}
\begin{split}
(\iFh - \iFc )&\big\{
 (\wDh-z)^{-1} \oplus \mathbf 0_h 
\big\}\Fc  \varphi    \\
&=
(\iFh - \iFc )\big\{
 (\wDc-z)^{-1}  \oplus \mathbf 0_h 
\big\}\Fc  \varphi    \\
&\qquad +
(\iFh - \iFc )\big\{
 (\wDh-z)^{-1} \oplus \mathbf 0_h 
- (\wDc-z)^{-1}  \big\}\Fc \varphi,   
 \end{split}
 \end{gather}
where, by  \cite[Lemma 3.5]{SU2023}, the $L^2$ norm of  the term (\ref{eq:differnc5})  tends 
 to $0$ as $h \to 0$.
Combining  all the arguments above, we infer that
\begin{align*} 
&\limsup_{h \to 0}
 \Vert
\big\{
(\Dh-z)^{-1}\oplus \mathbf 0_h  
- (\Dc -z)^{-1}  
\big\} \varphi  
 \Vert_{ L^2(\mathbb R^3)^4} \\
 &\quad \le
2 \limsup_{h \to 0}
\Vert  
\big\{
 (\wDh-z)^{-1} \oplus \mathbf 0_h  - (\wDc-z)^{-1}  
\big\}\Fc \varphi
\Vert_{ L^2(\Rthr)^4 },
\end{align*}

We finish the proof of statement (i) by showing that
\begin{gather}
\begin{split}\label{eq:differnc8}
\lim_{h \to 0}
\Vert  
\big\{
 (\wDh-z)^{-1} \oplus \mathbf 0_h   
 -(\wDc-z)^{-1}  
\big\}\Fc \varphi
\Vert_{ L^2(\Rthr)^4 }=0.
\end{split}
\end{gather}
The expression in (\ref{eq:differnc8})
is estimated by
\begin{align}
&
\lim_{h \to 0}
\Vert  
\big\{
 (\wDh-z)^{-1} \oplus \mathbf 0_h 
 - (\wDc-z)^{-1}  
\big\}
Q_{1\!/h}\Fc \varphi
\Vert_{ L^2(\Rthr)^4 }  \label{eq:differnc9}\\
& \qquad +
\lim_{h \to 0}
\Vert  
\big\{
(\wDc-z)^{-1}  
\big\}
Q_{1\!/h}^{\perp}\Fc \varphi
\Vert_{ L^2(\Rthr)^4 }.  \nonumber
\end{align}
The second term of (\ref{eq:differnc9}) 
equals $0$ since
$$ (\wDc(\xi)-z)^{-1} = (|\xi|^2 + m^2 -z^2)^{-1} (\wDc (\xi)+z).$$
The first term of (\ref{eq:differnc9}) 
is bounded by
\begin{align*} 
\lim_{h \to 0}
&
\Vert  
 (\wDh-z)^{-1} \big( \wDc -\wDh)  
 (\wDc-z)^{-1}  Q_{1\!/h}
\Fc \varphi
\Vert_{ L^2(\Rthr)^4 } \\
&\le
\frac{1}{|\mathfrak{Im}\,z |}  
\lim_{h \to 0} \!
\Big\{ 
\!\! \int_{\Rthr}  \! \Big|  
\big( \wDc(\xi) -\wDh(\xi)  \big) 
\chi_{\Ththr}(\xi) (\wDc(\xi)-z)^{-1}   
[\Fc \varphi] (\xi) \Big|_{{\mathbb C}^{4}}^2  \! d\xi
{\Big\}}^{1/2} .
\end{align*}
Note that 
 $\Fc \varphi$ 
on the right hand side of the above inequality
 belongs to $\mathcal S (\mathbb R^3)^4$.
In a similar manner to \cite[Lemma 4.2]{SU2023},
one can show that
\begin{align*}
&\big\Vert \big( \wDh (\xi) - \wDc (\xi) \big)\chi_{\Ththr}(\xi) \big\Vert_{\mathbf B({\mathbb C}^{4})} 
\le
\frac{h}{2}|\xi|^2.  
\end{align*}
Equation (\ref{eq:differnc8}) then follows by the Lebesgue dominated convergence theorem.

\bigskip
(ii)
Let
$u_h =(y_h,   0 ,   0,  0 )^T\in L^2(\mathbb R^3)^4$,
\begin{equation*}
y_h(x) = h^{\frac 3 4}\,e^{i\frac\pi{2 h}(x_1 - x_2)}\,e^{-h(x_1^2 + x_2^2 + x_3^2)} \quad (x \in \mathbb R^3),
\end{equation*}
for all $h > 0$;
then
$\|u_h\|_{L^2(\mathbb R^3)^4} = \left(\frac \pi 2 \right)^{\frac 3 4}$.

As the function $e^{-x^2/2}$ $(x \in \mathbb{R})$ is its own
Fourier transform, we find
for all $\xi \in \mathbb R^3$
\begin{equation*}
 (\mathop{\mathcal F} y_h)(\xi) 
 = \frac 1{(4 h)^{\frac 3 4}}\,e^{-\frac 1{4h} [(\xi_1 - \frac \pi{2h})^2 + (\xi_2 + \frac \pi{2h})^2 + \xi_3^2]}. 
\end{equation*}
Let $z \in \mathbb C \setminus \mathbb R$.
In the following, we show that $(\mathbb D_m - z)\,u_h$ tends to 0 as $h\rightarrow 0$ (step 1),
whereas the norm of $(\mathbb D_{m,h} \oplus \mathbf 0_h - z)^{-1}\,u_h$ remains
bounded from below by $|m-z|^{-1}$, up to an error that vanishes in the limit
(step 3). In step 2, we prepare application of the embedded discrete Dirac operator by calculating the projection $P_h y_h$, up to small error.

{\it Step 1.\/}
Applying the Fourier transformation to the Dirac operator (\ref{eq:Dcont}) 
and taking the matrix inverse, we find
\begin{align*}
(\mathbb D_m - z)^{-1} u_h  
&= \iFc \frac 1{m^2 -z^2 + \xi_1^2 + \xi_2^2 + \xi_3^2} \\
&\qquad\times\left(\begin{matrix}
 m + z & 0 & \xi_3 & \xi_1 -i\xi_2 \\
 0 & m + z & \xi_1 + i\xi_2 & -\xi_3 \\
 \xi_3 & \xi_1 -i\xi_2 & -m + z & 0 \\
 \xi_1 + i \xi_2 & -\xi_3 & 0 & -m + z
\end{matrix}\right)
\left(\begin{matrix} \Fc y_h \\ 0 \\ 0 \\ 0 \end{matrix}\right).
\end{align*}
As the Fourier transform is an isometry on $L^2(\mathbb R^3)$,
\begin{align*}
\|(\mathbb D_m - z)^{-1} u_h\|_{L^2(\mathbb R^3)^4}^2 
&
= \int_{\mathbb R^3} 
\frac {|m +z|^2 + \xi_1^2 + \xi_2^2 + \xi_3^2}{|m^2 - z^2 + \xi_1^2 + \xi_2^2 + \xi_3^2 \,|^2} \\
&\qquad\times\frac 1{(4h)^{\frac 3 2}}\,e^{-[(\xi_1 - \frac \pi{2h})^2 + (\xi_2 + \frac \pi{2h})^2 + \xi_3^2]/2h}\,d\xi.
\end{align*}
Now we observe that there is a constant $C_z > 0$ (which may depend on $z$)
such that
\[
 \frac{|m + z|^2 + \xi_1^2 + \xi_2^2 + \xi_3^2}{|m^2 - z^2 + \xi_1^2 + \xi_2^2 + \xi_3^2|} \le C_z \qquad (\xi \in \mathbb{R}^3),
\]
since the denominator is bounded below by a positive constant and the fraction tends to 1 as $|\xi|\rightarrow\infty$.
Hence
\begin{align*}
\|(\mathbb D_m - z)^{-1} u_h\|_{L^2(\mathbb R^3)^4}^2
&\le C_z \int_{\mathbb R^3}
 \frac{e^{-[(\xi_1  - \frac \pi{2h})^2 + (\xi_2 + \frac \pi{2h})^2 + \xi_3^2]/2h}}{(4h)^{\frac 3 2}\,|m^2 - z^2 + \xi_1^2 + \xi_2^2 + \xi_3 ^2|}\,d\xi
\\
&= \frac{C_z}8  \int_{\mathbb R^3}
\frac {e^{-(\zeta_1^2 + \zeta_2^2 + \zeta_3^2)/2}} {|m^2 \!- \! z^2 \!+\! ( \!\sqrt h \zeta_1 \!+\! \frac \pi{2h})^2 \!+\! (\!\sqrt h \zeta_2 \!-\! \frac \pi{2h})^2 \!+\! h \zeta_3^2|}
\,d\zeta.
\end{align*}
The last integral converges to 0 as $h \rightarrow 0$ by
dominated convergence, as the integrand tends to 0 pointwise and
can be bounded above by an $h$-independent multiple of $e^{-(\zeta_1^2 + \zeta_2^2 + \zeta_3^2)/2}$.
We have thus established that
\begin{equation*}
\lim_{h\rightarrow 0} \|(\mathbb D_m - z)^{-1} u_h\|_{L^2(\mathbb R^3)^4} = 0.
\end{equation*}

{\it Step 2.\/}
Before we can apply the discrete Dirac operator (\ref{eq:Dd}), we need to project
$u_h$ into the subspace
$L^2(\mathbb Z_h^3)^4$.
By 
equation (\ref{eq:ogp}), the projection is given by
$P_h y_h = \sum_{n\in\mathbb Z^3} \widetilde{y}_h (hn) \charnh,  $ 
where
\begin{equation*}
\widetilde{y}_h(hn) 
= h^{\frac 3 4} \left(\int_{n_1}^{n_1+1} e^{i\frac \pi 2 t}\,e^{-h^3 t^2}\,dt\right) 
\left(\int_{n_2}^{n_2+1} e^{-i\frac \pi 2 t}\,e^{-h^3 t^2}\,dt\right) 
\!\! \left(\int_{n_3}^{n_3+1} e^{-h^3 t^2}\,dt\right)
\end{equation*}
for all $n \in \mathbb Z^3$.
By an integration by parts, we find for all $k\in\mathbb Z$
\begin{gather}
\begin{split}\label{eq:Kplusminus}
 \int_{k}^{{k}+1} e^{\pm i\frac\pi2 t} e^{-h^3 t^2}\,dt   = \frac 2\pi (1 \pm i)\, e^{\pm i\frac\pi 2 {k}} e^{-h^3{k}^2} + \frac {4 h^3}\pi K^{\pm}_{{{k}},h}
\end{split}
\end{gather}
with
\begin{align*}
 K^{\pm}_{k,h} &= \frac{e^{\pm i\frac\pi 2 k}}{2 h^3}\,
 (e^{-h^3(k+1)^2} - e^{-h^3 k^2}) 
\mp i 
 \int_k^{k+1} e^{\pm i\frac\pi 2 t}\,t\,e^{-h^3 t^2}\,dt
\\
 &= \int_k^{k+1} \!\! e^{\pm i\frac\pi 2 k}\,t\,e^{-h^3 t^2}dt
\mp i  \!\! \int_k^{k+1}  \!\! e^{\pm i\frac\pi 2 t}\,t\,e^{-h^3 t^2}\,dt.
\end{align*}
Noting that
$x e^{-x^2/2} \le \frac 1 {\sqrt e}$ $(x \ge 0)$, we obtain the estimate
\begin{eqnarray}
\sum_{k\in\mathbb Z} |h^{\frac 3 2} K^\pm_{k,h}|^2   \label{eq:Kpmest}
&\le \displaystyle{\sum_{k\in\mathbb Z}} \left(2\int_k^{k+1} h^{\frac 3 2}\,|t|\,e^{-(h^{\frac 3 2}|t|)^2/2}\,e^{-h^3 t^2/2}\,dt \right)^2 \\
&\hspace{-68pt}\le \dfrac 4 e \displaystyle{\sum_{k\in\mathbb Z} }\left(\int_k^{k+1} e^{-h^3 t^2/2}\,dt\right)^2  \nonumber\\
&\hspace{-8pt}\le \dfrac 8 e  \displaystyle{\sum_{k=0}^\infty }e^{-h^3 {k^2}}
= \dfrac{4\sqrt\pi}{e h^{\frac 3 2}} + O(1) \qquad (h \rightarrow 0)
\nonumber
\end{eqnarray}
by (\ref{eq:asy1}).
Similarly,
\begin{equation}\label{eq:Kzero}
 \int_k^{k+1} e^{-h^3 t^2}\,d t = e^{-h^3 k^2} + 2 h^3 K^0_{k,h}
\end{equation}
with
\begin{align*}
K^0_{k,h} &= \frac 1{2 h^3} \int_k^{k+1} (e^{-h^3 t^2} - e^{-h^3 k^2})\,dt
= - \int_k^{k+1} \int_k^t s\,e^{-h^3 s^2}\,ds\,dt
\\
&= - \int_k^{k+1} (k+1-s)\,s\,e^{-h^3 s^2}\,ds.
\end{align*}
and we have the estimate
\begin{eqnarray}
\displaystyle{\sum_{k\in\mathbb Z}} |h^{\frac 3 2} K^0_{k,h}|^2 
& \le 
\displaystyle{\sum_{k\in\mathbb Z}} \left(\int_k^{k+1}\hspace{-5pt} h^{\frac 3 2}\,|s|\,e^{-(h^{\frac 3 2} |s|)^2/2}\,e^{-h^3 
{ s^2}/2}\,ds \right)^2
\label{eq:Kzest}
\\
&\le \dfrac 1 e 
\displaystyle{\sum_{k\in\mathbb Z}} \left(\int_k^{k+1}\hspace{-5pt} e^{-h^3 s^2/2}\,ds\right)^2 
 \le \frac{\sqrt\pi}{e h^{\frac 3 2}} + O(1) 
 \nonumber
\end{eqnarray}
$(h\rightarrow 0)$.
Using equations (\ref{eq:Kplusminus}) and (\ref{eq:Kzero}), we can write
$\widetilde y_h(h n)$ in the form
\begin{align*}
\widetilde y_h(h n) 
&= h^{\frac 3 4} \left(\frac 2\pi\,(1 + i)\,e^{i\frac\pi 2 n_1}e^{-h^3 n_1^2} + \frac {4 h^3}\pi\,
K^+_{{n_1},h}\right)
\\ 
&\qquad
\times \left(\frac 2\pi\,(1-i)\,e^{-i\frac\pi 2 n_2}e^{-h^3 n_2^2} 
+ \frac {4 h^3}\pi\,
K^-_{{n_2},h}\right) 
\left(e^{h^3 n_3^2} + 2 h^3 K^0_{{n_3},h}\right) 
 \\
&
= \frac 8{\pi^2} y_h(n h) +  \sum_{j=1}^7 R_{j,n,h}
\end{align*}
where
\begin{align*}
 R_{1,n,h} &= h^{3+\frac 3 4}\,\frac{16}{\pi^2}\,e^{i\frac\pi 2\,(n_1-n_2)} e^{-h^3(n_1^2 + n_2^2)}\,K^0_{n_3,h},
\\
 R_{2,n,h} &= h^{3 + \frac 3 4}\,\frac 8{\pi^2}\,(1+i)\,e^{i\frac\pi 2\,n_1} e^{-h^3(n_1^2 + n_3^2)}\,K^-_{n_2,h},
\\
 R_{3,n,h} &= h^{3 + \frac 3 4}\,\frac 8{\pi^2}\,(1-i)\,e^{-i\frac\pi 2\,n_2} e^{-h^3(n_2^2 + n_3^2)}\,K^+_{n_1,h},
\\
 R_{4,n,h} &= h^{6+\frac 3 4}\,\frac{16}{\pi^2}\,(1+i)\,e^{i\frac\pi 2 n_1} e^{-h^3 n_1^2}\,K^-_{n_2,h}\,K^0_{n_3,h},
\\
 R_{5,n,h} &= h^{6+\frac 3 4}\,\frac{16}{\pi^2}\,(1-i)\,e^{-i\frac\pi 2 n_2} e^{-h^3 n_2^2}\,K^+_{n_1,h}\,K^0_{n_3,h},
\\
 R_{6,n,h} &= h^{6+\frac 3 4}\,\frac{16}{\pi^2}\,e^{-h^3 n_3^2}\,K^+_{n_1,h}\,K^-_{n_2,h},
\\
 R_{7,n,h} &= h^{9+\frac 3 4}\,\frac{32}{\pi^2}\,K^+_{n_1,h}\,K^-_{n_2,h}\,K^0_{n_3,h}.
\end{align*}
With the notation
\begin{equation*}
 R_h^{(j)} := \sum_{n\in\mathbb Z^3} R_{j,n,h}\,\chi_{I_{n,h}}   \qquad   (j \in \{1, \, \cdots, \, 7 \}),
\end{equation*}
we  embed the remainder term into $L^2(\mathbb Z_h^3)$ by defining
 $R_h 
   = \sum_{j=1}^7 R_h^{(j)}$
and also set
$(y_h)_h = J_h y_h$;
then
\begin{equation}\label{eq:Phyh}
 \mathop{P_h} y_h = \frac 8{\pi^2}\,(y_h)_h +  R_h.
\end{equation}
Each $R_h^{(j)}$, where $j\in\{1,\dots,7\}$, can be estimated as follows.
\begin{align*}
\| R_h^{(1)}  \|_{L^2(\mathbb Z_h^3)}^2
&= h^{3+\frac 9 2}\,\frac{256}{\pi^4} \left(\sum_{n_1\in\mathbb Z} e^{-2h^3 n_1^2}\right) 
\left(\sum_{n_2\in\mathbb Z} e^{-2h^3 n_2^2}\right) \!\! \left(\sum_{n_3\in\mathbb Z} |h^{\frac 3 2} K^0_{n_3,h}|^2\right)
\\
&\le h^{3+\frac 9 2}\,\frac{256}{\pi^4} \left(\sqrt{\frac \pi{2 h^3}} + O(1) \right)^2 \left(\frac{\sqrt\pi}{e h^{\frac 3 2}} + O(1)\right)
\\
&= \frac{128}{\sqrt\pi\,e}\,h^3 + O(h^{\frac 9 2}),
\end{align*}
where we used the first asymptotic formula in Lemma \ref{lem:esums} and the estimate (\ref{eq:Kzest}).
Analogously,
\begin{align*}
\| R_h^{(2)} \|_{L^2(\mathbb Z_h^3)}^2 
&\le h^{3+3+\frac 3 2}\,\frac{128}{\pi^4} \left(\sqrt{\frac \pi{2h^3}} + O(1)\right)^2 \!\! \left(\frac{4\sqrt\pi}{e\,h^{\frac 3 2}} + O(1)\right)
\\
&= \frac{256}{\pi^{\frac 5 2}e}\,h^3 + O(h^{\frac 9 2})
\end{align*}
using the estimate (\ref{eq:Kpmest}); the same estimate holds for
$\| R_h^{(3)} \|_{L^2(\mathbb Z_h^3)}^2$.
Further,
\begin{align*}
\|R_h^{(4)} \|^{2}_{L^2(\mathbb Z_h^3)}
&\le h^{3+6+\frac 3 2}\,\frac{512}{\pi^4}\left(\sqrt{\frac\pi{2 h^3}} + O(1)\right) 
\left(\frac{4\sqrt\pi}{e\,h^{\frac 3 2}} + O(1)\right) \!\! \left(\frac{\sqrt\pi}{e\,h^{\frac 3 2}} + O(1)\right)
\\
&\quad
= \frac{1024}{\sqrt 2\,\pi^{\frac 5 2}e^2}\,h^6 + O(h^{\frac{15}2}),
\end{align*}
and the same estimate holds for
$\| R_h^{(5)}\|_{L^2(\mathbb Z_h^3)}^2$.
We find
\begin{align*}
\|R_h^{(6)}\|^2_{L^2(\mathbb Z_h^3 ) } & 
\le h^{3+6+\frac 3 2}\,\frac{256}{\pi^4} \left(\sqrt{\frac\pi{2h^3}} + O(1)\right) 
\left(\frac{4\sqrt\pi}{e\,h^{\frac 3 2}} + O(1)\right)^2 \\
&= \frac{4096}{\sqrt 2\,\pi^{\frac 5 2} e^2}\,h^6 + O(h^{\frac{15}2})
\end{align*}
and finally
\begin{align*}
\| R_h^{(7)}\|_{L^2(\mathbb Z_h^3)}^2 & 
\le h^{3+9+\frac 3 2} \frac{1024}{\pi^4}\left(\frac{4\sqrt\pi}{e\,h^\frac 3 2} + O(1)\right)^2 \!\! \left(\frac{\sqrt\pi}{e\,h^\frac 3 2} + O(1)\right) 
\\
&= \frac{16384}{\pi^{\frac 5 2} e^3}\,h^9 + O(h^{\frac{21}2}).
\end{align*}
In total, this gives
$\|R_h\|_{L^2(\mathbb Z_h^3)} = O(h^\frac 3 2)$ as $h \rightarrow 0$.
By (\ref{eq:asy1}), the first term in (\ref{eq:Phyh}) satisfies
\begin{align*}
\|\frac 8{\pi^2} (y_h)_h\|_{L^2(\mathbb Z_h^3)}^2 
&
= h^{3+\frac 3 2}\,\frac{64}{\pi^4} \left(\sqrt{\frac\pi{2h^3}} + O(1)\right)^3,
\end{align*}
so
$\|\frac 8{\pi^2} (y_h)_h\|_{L^2(\mathbb Z_h^3)} = \frac 8{\pi^\frac 5 4 2^\frac 3 4} + O(h^\frac 3 2)$.

{\it Step 3.\/}
We now apply the discrete Dirac operator,
\begin{equation}\label{eq:dDy}
(\mathbb D_{m,h} - z) \left(
\begin{matrix} (y_h)_h \\ 0 \\ 0 \\ 0 \end{matrix}\right) \\
= \left(\begin{matrix}
 (m-z)\,(y_h)_h \\ 0 \\ -i\partial_{3,h} (y_h)_h
 \\ (-i\partial_{1,h} + \partial_{2,h}) (y_h)_h \end{matrix}\right).
\end{equation}
Here
\begin{align*}
D_{1,h}(nh) &:= -i\partial_{3,h} (y_h)_h(nh)  \qquad (n\in\mathbb Z^3)\\
&
= -i h^{-\frac 1 4} e^{ i  \frac\pi 2(n_1 - n_2)} e^{-h^3(n_1^2 + n_2^2 + n_3^2)} 
\left(e^{-h^3(2 n_3 + 1)} - 1\right);
\end{align*}
noting that
\[ \left| e^{-h^3(2k +1)} - 1 \right|  \!=\! \left|-h^3  \!\!\!
 \int_0^{2k +1}  \!\! e^{-h^3 t}\,dt\right|
\!\le \! (2 k  + 1)\,h^3
\]
and
$(2k  + 1)^2 \le 1 + 8 k^2$
for $ k \in\mathbb N_0$, we find, using both asymptotic formulae in Lemma \ref{lem:esums},
\begin{align*}
\|D_{1,h}\|_{L^2(\mathbb Z_h^3)}^2 &\le h^{3-\frac 1 2} \left(\sum_{n_1\in\mathbb Z} e^{-2h^3 n_1^2} \!\right) \!\! \left(\sum_{n_2\in\mathbb Z} e^{-2h^3 n_2^2} \!\right) 
2 \left(\sum_{n_3\in\mathbb N_0} e^{-2h^3 n_3^2} (2n_3+1)^2 h^6\!\!\right)
\\
&
\le h^{3-\frac 1 2 + 6} \left( \frac\pi{2 h^3} + O(\frac 1{h^\frac 3 2})\right)
\left(\frac{\sqrt{2\pi}}{h^\frac 9 2} + O(\frac 1{h^3})\right)
\\
&
= \frac{\sqrt\pi^3}{\sqrt 2}\,h + O(h^\frac 5 2)
\qquad (h\rightarrow 0).
\end{align*}
Hence
$\|D_{1,h}\|_{L^2(\mathbb Z_h^3)} \le \left(\frac{\pi^3}2\right)^\frac 1 4\sqrt h + O(h^2)$. 
Similarly, we find for
\begin{align*}
D_{2,h}(hn) &:= (-i\partial_{1,h} + \partial_{2,h}) (y_h)_h (hn )\\
&= (-i\partial_{1,h} + \partial_{2,h}) h^\frac 3 4 \,e^{i\frac \pi 2 (n_1 - n_2)} e^{-h^3(n_1^2+n_2^2+n_3^2)}
\\
&
= h^{-\frac 1 4} e^{i\frac \pi 2 (n_1 - n_2)} e^{-h^3(n_1^2 + n_2^2 + n_3^2)} 
\left((e^{-h^3(2 n_1 + 1)} - 1) -i\,(e^{-h^3(2 n_2 + 1)} - 1)\right)
\end{align*}
the norm estimate
$\|D_{2,h}\|_{L^2(\mathbb Z_h^3)} \le (2\pi)^{\frac 34}\sqrt{h}  + O(h^2)$.

To complete the proof, we first note that
\begin{equation*}
(\mathbb D_{m,h} \oplus \mathbf 0_h - z)^{-1} = (\mathbb D_{m,h} - z)^{-1} \oplus \left(-\frac 1 z\right).
\end{equation*}
By equations (\ref{eq:Phyh}) and (\ref{eq:dDy}),
\begin{equation*} 
(\mathbb D_{m,h} - z) P_h u_h 
= \frac 8{\pi^2} (m-z)\,(u_h)_h
+ \frac 8{\pi^2}
  \left(\begin{matrix} 0 \\ 0 \\ D_{1,h} \\ D_{2,h} \end{matrix}\right)  
+ (\mathbb D_{m,h} - z) 
\left(\begin{matrix} R_h \\ 0 \\ 0 \\ 0 \end{matrix}\right), \nonumber
\end{equation*}
and, using (\ref{eq:Phyh}) again in the first term on the right hand side and
applying
$\frac{1}{m-z}(\mathbb D_{m,h} - z)^{-1}$
on both sides of the equation, we find
\begin{align*}
(\mathbb D_{m,h} - z)^{-1} P_h u_h
&= \frac 1{m-z}\,P_h u_h
 - \frac 8{\pi^2}\, \frac 1{m-z}\,(\mathbb D_{m,h} - z)^{-1}  \!\!
\left(\!\begin{matrix} 0 \\ 0 \\ D_{1,h} \\ D_{2,h}  \end{matrix} \!\right) 
\\
&\qquad
- \frac 1{m-z} \!
\left(\begin{matrix} R_h \\ 0 \\ 0 \\ 0 \end{matrix}\right)
 + (\mathbb D_{m,h} - z)^{-1} \left(\begin{matrix} R_h \\ 0 \\ 0 \\ 0 \end{matrix}\right).
\end{align*}
Using the estimate (\ref{eq:resest}),
we hence obtain
\begin{align*}
&\left\|(\mathbb D_{m,h} - z)^{-1} P_h u_h - 
\frac 1{m-z}\,P_h u_h\right\|_{L^2(\mathbb R^3)^4}
\\
&\qquad \le \frac{8
\sqrt{\|D_{1,h}\|_{L^2(\mathbb Z_h^3)}^2 + \|D_{2,h}\|_{L^2(\mathbb Z_h^3)}^2}
}{|m-z|\,  |\mathfrak{Im}\,z | \,\pi^2} 
 + \left(\frac 1{|m-z|} + \frac{1}{|\mathfrak{Im}\,z | } \right) 
 \| R_h \|_{L^2(\mathbb Z_h^3)}
\rightarrow 0
\end{align*}
as $h\rightarrow 0$ by our remainder term estimates.
Thus
\begin{align*}
\|u_h&\|_{L^2(\mathbb R^3)^4}
\|(\mathbb D_{m,h} \oplus \mathbf 0_h - z)^{-1}
 - (\mathbb D_m - z)^{-1}\|_{\mathbf B}
\\
&\ge \|(\mathbb D_{m,h} - z)^{-1} P_h u_h  -\frac 1 z\,(1 - P_h) u_h\|_{L^2(\mathbb R^3)^4} 
- \|(\mathbb D_m - z)^{-1} u_h\|_{L^2(\mathbb R^3)^4}
\\
&= \Big\{\|(\mathbb D_{m_h} - z)^{-1} P_h u_h\|_{L^2(\mathbb R^3)^4}^2 
+ \frac 1{|z|^2} \, \| (1 - P_h) u_h\|_{L^2(\mathbb R^3)^4}^2\Big\}^{1/2} + o(1)
\\
&\ge \|(\mathbb D_{m,h} - z)^{-1} \mathop{P_h} u_h\|_{L^2(\mathbb R^3)^4} + o(1)
\\
&= \left\|\frac 1{m-z} \mathop{P_h} u_h\right\|_{L^2(\mathbb R^3)^4} + o(1)
= \frac 1{|m-z|}\,\|u_h\|_{L^2(\mathbb R^3)^4} + o(1) \qquad (h\rightarrow 0),
\end{align*}
where we used in the last step the strong convergence of $\mathop{P_h}$ to the
identity operator., see \cite[Lemma 3.3]{SU2023}.
As $\|u_h\|_{L^2(\mathbb R^3)^4}$ is independent of $h$, it follows that
\begin{align*}
\liminf_{h\to 0}
\Vert 
({\mathbb D}_{m,h} \oplus \mathbf 0_h - z )^{-1}  
-  ({\mathbb D}_{m} - z)^{-1}
 \, \Vert_{\mathbf B}
 \ge \frac 1 {|m - z|}.
\end{align*}
The second lower bound in statement (ii) follows analogously starting from
$u_h = (0, 0, y_h, 0)^T$.
\hfill$\blacksquare$\break

\bigskip\noindent
{\bf Acknowledgement.}
The authors thank the anonymous reviewer for helpful and constructive comments.

  \end{document}